\documentclass[journal,doublecolumn]{IEEEtran}
\usepackage{amsmath,amssymb,verbatim,cite,amsopn,graphicx,citesort,url,color,mathtools,epsfig,epstopdf,enumerate,balance}
\usepackage{float}
\usepackage{amsopn,graphicx,citesort,url,color}
\usepackage{algorithm}
\usepackage{algorithmic}
\usepackage{multicol}
\usepackage{epsfig}
\usepackage{epstopdf}
\usepackage{subcaption}
\usepackage[font={small}]{caption}

\usepackage{lipsum}
\newcommand{\M}{M}
\newcommand{\X}{X}
\newcommand{\Z}{Z}
\newcommand{\Y}{Y}
\newcommand{\vr}{\gamma}
\newcommand{\vrba}{\bar{\gamma_b}}
\newcommand{\vrea}{\bar{\gamma_e}}

\newcommand{\Dta}{\bar{\Delta}}

\newcommand{\pout}{p_\text{out}}

\newcommand{\ptx}{p_\text{tx}}
\newcommand{\Rsmin}{R_{s,\min}}
\newcommand{\Rsmax}{R_{s,\max}}
\newtheorem{Proposition}{Proposition}

\newtheorem{Remark}{Remark}

\begin{document}

\title{On Secrecy Metrics for Physical Layer Security over Quasi-Static Fading Channels}

\author{
Biao~He,~\IEEEmembership{Member,~IEEE,}
Xiangyun~Zhou,~\IEEEmembership{Member,~IEEE,}
A.~Lee~Swindlehurst,~\IEEEmembership{Fellow,~IEEE}
\thanks{This work was supported by the Australian Research Council under Discovery Project Grant DP150103905. This work was presented in part at the 2014 IEEE Global Communications Conference (GLOBECOM)~\cite{He_14_New}.}
\thanks{B. He is with Department of Electronic and Computer Engineering,
Hong Kong University of Science and Technology, Hong Kong (email:
eebiaohe@ust.hk).}
\thanks{X. Zhou is with the Research School of Engineering, The Australian National University, Canberra, ACT 2601, Australia (e-mail: xiangyun.zhou@anu.edu.au).}
\thanks{A. L. Swindlehurst is with the Center for Pervasive Communications and Computing, Department of Electrical Engineering and Computer Science, University of California, Irvine, CA 92697, USA (e-mail:swindle@uci.edu).}
}

\maketitle

\begin{abstract}
Theoretical studies on physical layer security often adopt the secrecy outage probability as the performance metric for wireless communications over quasi-static fading channels. The secrecy outage probability has two limitations from a practical point of view: a) it does not give any insight into the eavesdropper's decodability of confidential messages; b) it cannot characterize the amount of information leakage to the eavesdropper when an outage occurs.
Motivated by the limitations of the secrecy outage probability, we propose three new secrecy metrics for secure transmissions over quasi-static fading channels. The first metric establishes a link between the concept of secrecy outage and the decodability of messages at the eavesdropper. The second metric provides an error-probability-based secrecy metric which is typically used for the practical implementation of secure wireless systems. The third metric characterizes how much or how fast the confidential information is leaked to the eavesdropper. We show that the proposed secrecy metrics collectively give a more comprehensive understanding of physical layer security over fading channels and enable one to appropriately design secure communication systems with different views on how secrecy is measured.
\end{abstract}


\section{Introduction}\label{sec:Intro}

\subsection{Background and Motivation}\label{Sec:IA}
\IEEEPARstart{A}{n} unprecedented amount of private and sensitive information is transmitted over wireless channels as a result of the ubiquitous wireless devices adopted in modern life.
Security issues associated with wireless communications consequently have become critical due to the unchangeable open nature of the wireless medium.
As a complement to traditional cryptographic techniques, physical layer security has been proposed for ensuring secure wireless communications by exploiting the characteristics of
wireless channels~\cite{Bloch_11,Zhou_13_Physical}.
Shannon~\cite{Shannon_49} introduced the notion of information-theoretic secrecy, which does not rely on assumptions about the computational abilities of the eavesdropper.
Classical information-theoretic secrecy\footnote{In this paper, we use the term ``classical information-theoretic secrecy" to refer to Shannon's perfect secrecy,  strong secrecy, and weak secrecy, which will be described later in Section~\ref{sec:PerfectSecrecy}.} requires that the amount of information leakage to the eavesdropper vanishes. It guarantees that the eavesdropper's optimal attack is to guess the message at random, and hence the eavesdropper's decoding error probability, $P_e$, asymptotically goes to 1. In his seminal work~\cite{wyner_75}, Wyner introduced the wiretap channel, and addressed the tradeoff between the information rate achieved by the intended receiver and the level of ignorance at the eavesdropper. This result was later extended to the broadcast channel with confidential messages \cite{csiszar_78} and the Gaussian wiretap channel~\cite{Cheong_78}.

More recently, physical layer security over wireless fading channels has been extensively studied, e.g., \cite{Gopala_08,Liang_08,Khisti_08,Bloch_08,Zhou_11}. In particular, practical scenarios involving imperfect or no knowledge about the eavesdropper's instantaneous channel state information (CSI) has drawn an increasing amount of attention, e.g., see~\cite{He_13_3} and references therein.
The secrecy performance in such scenarios is often characterized by either ergodic secrecy capacity~\cite{Gopala_08} or secrecy outage probability~\cite{Bloch_08,Zhou_11}.
For a system in which the encoded messages can span sufficient channel realizations to capture the ergodic features of the fading channel, the ergodic secrecy capacity characterizes the capacity limit subject to the constraint of classical information-theoretic secrecy.
For transmission over quasi-static fading channels where classical information-theoretic secrecy is not always achievable, the (classical) secrecy outage probability measures the probability of failing to achieve classical information-theoretic secrecy.
With either the ergodic secrecy capacity or the secrecy outage probability as the secrecy metric, many researchers have studied secure transmission designs and/or secrecy enhancements, e.g.,~\cite{Mukherjee_11,Yang_13_Transmit,He_13_2,Chen_11_Joint,Wang_14_Secure}. 


Classical secrecy outage probability has two major limitations in evaluating the secrecy performance of wireless systems.
\begin{enumerate}[a)]
    \item
    Classical secrecy outage probability does not give any insight into the eavesdropper's ability to decode the confidential messages.
    The eavesdropper's decodability is an intuitive measure of security in real-world communication systems when
    classical information-theoretic secrecy is not always achievable, and error-probability-based secrecy metrics are often adopted to quantify secrecy performance in the literature,
    e.g.,~\cite{Belfiore_13_AnError,Karpuk_15_probability,Belfiore_10_SecrecyGain} focusing on infinite-length code design,~\cite{Baldi_15_Performance,Klinc_11_LDPC,Baldi_12_Coding} investigating finite-length coding schemes,~\cite{Soosahabi_12_Scalable} utilizing probabilistic ciphering,~\cite{Khan_15_Rethinking} investigating secure network coding, and~\cite{Barcelo_14_amplify} studying secrecy with compressive sensing.
    A general secrecy requirement for the eavesdropper's decoding error probability can be given as $P_e\ge\epsilon$, where $0<\epsilon\le1$ denotes the minimum acceptable value of $P_e$. In contrast, classical secrecy outage probability reflects only an extremely stringent requirement on $P_e$ for $\epsilon\rightarrow1$, i.e., requiring $\epsilon\rightarrow1$, since classical information-theoretic secrecy guarantees $P_e\rightarrow1$.
\item The amount of information leakage to the eavesdropper cannot be characterized. When classical information-theoretic secrecy is not achievable, some information will be leaked to the eavesdropper.
    Different secure transmission designs that lead to the same secrecy outage probability may actually result in very different amounts of information leakage.
    Consequently, it is important to know how much or how fast the confidential information is leaked to the eavesdropper to obtain a finer view of the secrecy performance. However, the classical outage-based approach is not able to evaluate the amount of information leakage when a secrecy outage occurs.
\end{enumerate}
It is worth mentioning that, apart from the two above mentioned limitations, the classical secrecy outage probability also has a severe limitation in evaluating the secrecy performance of systems with finite-length coding schemes. Since classical information-theoretic secrecy cannot be achieved by any coding scheme with a finite-length codeword, the classical secrecy outage probability based on the classical information-theoretic secrecy cannot be adopted in the studies focusing on finite-length coding schemes. Thus, it is of significant importance to examine secrecy metrics specifically for wireless systems with finite-length codes, although such a study is beyond the scope of this paper.

\subsection{Our Approach and Contribution}
As previously discussed, the classical information-theoretic secrecy is not always achievable for transmissions over quasi-static fading channels, and we cannot ensure that the eavesdropper's decoding error probability always goes to 1. The classical secrecy outage probability, which is the secrecy metric for quasi-static fading channels, in fact has limitations in evaluating the secrecy performance of wireless systems. This motivates us to propose new secrecy metrics for wireless transmissions focusing on quasi-static fading channels in this paper.
The classical secrecy outage probability is based on the concept of classical information-theoretic secrecy. On the other hand, our proposed secrecy metrics are based on another regime of interest in physical layer security, namely the \emph{partial} secrecy regime.
The partial secrecy of a system is often evaluated using the equivocation, which reflects the level at which the eavesdropper is confused.
The study of equivocation for secrecy can be found as early as Wyner's pioneering work for the wiretap channel~\cite{wyner_75}. Similarly, Csisz\'{a}r and K\"{o}rner~\cite{csiszar_78} used the normalized equivocation to quantify partial secrecy for the broadcast channel with confidential information.
Importantly, the equivocation is closely related to the decoding error probability~\cite{Cover_06,Feder_94,wyner_75}. Therefore, evaluating the secrecy performance on the basis of equivocation can reflect the decodability of confidential messages at the eavesdropper.




Specifically, we propose three new secrecy metrics:
\begin{enumerate}
 \item Extended from the classical definition of secrecy outage, a generalized formulation of secrecy outage probability is proposed. The generalized secrecy outage probability takes into account the level of secrecy measured by equivocation, and hence establishes a link between the concept of secrecy outage and the decodability of messages at the eavesdropper.
 \item An asymptotic lower bound on the eavesdropper's decoding error probability is proposed. This proposed metric provides a \textsl{direct} link to error-probability-based secrecy metrics that are often used for the practical implementation of security in wireless systems operating over fading channels.
 \item A metric evaluating the average information leakage rate is proposed.
 This proposed secrecy metric gives an answer to the important question of how much or how fast the confidential information is leaked to the eavesdropper when classical information-theoretic secrecy is not achieved. 
\end{enumerate}
We note that both the generalized secrecy outage probability and the asymptotic lower bound on the eavesdropper's decoding error probability give insights into the eavesdropper's ability to decode the confidential messages. In comparing these two metrics, we highlight that the asymptotic lower bound on the eavesdropper's decoding error probability provides a more direct bridge to the error-probability-based secrecy metrics.
Although the eavesdropper's decoding error probability cannot be exactly characterized, the asymptotic lower bound gives a worst-case estimation of the eavesdropper's decodability.
On the other hand, the generalized secrecy outage probability is extended from the classical secrecy outage probability. Hence, existing studies on secrecy outage probability can be easily extended to the generalized secrecy outage probability.

To illustrate the use of the newly proposed secrecy metrics, we evaluate the secrecy performance of an example wireless system with fixed-rate wiretap codes.
We show that the proposed secrecy metrics can provide a more comprehensive and in-depth understanding of the secrecy performance over fading channels.
Moreover, we investigate the impact of the new secrecy metrics on the transmission design. We find that the newly proposed secrecy metrics lead to very different optimal design parameters that optimize the secrecy performance of the system, compared with the optimal design minimizing the classical secrecy outage probability. We also find that applying the optimal design that minimizes the secrecy outage probability can result in a large secrecy loss, if the actual system requires a low decodability at the eavesdropper and/or a low information leakage rate.
%
%

It is worth mentioning that this work is solely motivated by the limitations of the classical secrecy outage probability from a more practical point of view. Our proposed new secrecy metrics based on the concept of partial secrecy do not imply that the secrecy metrics based on classical information-theoretic secrecy are inappropriate from the information-theoretic perspective. We acknowledge the importance of requiring classical information-theoretic secrecy for research on information-theoretic security.
Meanwhile, we notice the large gap between the requirement of information-theoretic security and the condition of practical secrecy. We hope that the newly proposed secrecy metrics can enable contributions that bridge the gap between theory and practice in physical layer security.

The remainder of the paper is organized as follows.
Section~\ref{sec:Intro-back} provides background information on classical information-theoretic secrecy and partial secrecy.
Section~\ref{sec:content} introduces the three new secrecy metrics for wireless transmissions over fading channels. 
Section~\ref{sec:example} illustrates the use of the newly proposed metrics by evaluating the secrecy performance of an example wireless system with fixed-rate wiretap codes. Section~\ref{sec:Design} demonstrates the impact of the new secrecy metrics on system design, and finally Section~\ref{sec:Conc} concludes the paper.

\section{Preliminaries}\label{sec:Intro-back}
\begin{figure}[!htb]
\includegraphics[width=\columnwidth]{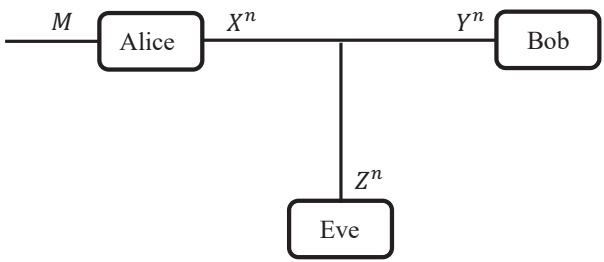}
\vspace{-3mm}
\caption{Basic wiretap channel.}
\vspace{-0mm}  \label{fig:BasicModel}
\end{figure}

Consider the basic wiretap-channel system shown in Figure~\ref{fig:BasicModel}. A transmitter, Alice, sends confidential information, $\M$, to an intended receiver, Bob, in the presence of an eavesdropper, Eve.
The source is stationary and ergodic. The confidential information, $\M$, is encoded into a $n$-vector $\X^n$.
The received vectors at Bob and Eve are  denoted by $\Y^n$ and $\Z^n$, respectively.
The entropy of the source information and the residual uncertainty for the message at the eavesdropper are denoted by $H(\M)$ and $H(\M \mid \Z^n)$, respectively.

\subsection{Classical Information-Theoretic Secrecy}\label{sec:PerfectSecrecy}
As mentioned before, classical information-theoretic secrecy implies that the amount of information leakage to the eavesdropper vanishes, and guarantees that the eavesdropper's optimal attack is to guess the message at random.
From Shannon's definition, perfect secrecy requires statistical independence between the original message and Eve's observation, which is given by
\begin{equation}\label{}
 H(\M\mid \Z^n)=H(\M) ~ \text{or, equivalently}, ~ I(\M; \Z^n)=0.
\end{equation}
Since Shannon's definition of perfect secrecy is not convenient to be used for further analysis, current research often investigates strong secrecy or weak secrecy. Strong secrecy requires asymptotic statistical independence of the message and Eve's observation as the codeword length goes to infinity, i.e., $\lim_{n\rightarrow\infty}I(\M; \Z^n)=0$. Weak secrecy requires that the rate of information leaked to the eavesdropper vanishes, i.e., $\lim_{n\rightarrow\infty}\frac{1}{n}I(\M; \Z^n)=0$.
Since strong secrecy, weak secrecy and Shannon's perfect secrecy all belong to the classical information-theoretic secrecy regime, for simplicity we use the term ``classical information-theoretic secrecy" to refer to such a regime in this paper. For simplicity, we also do not explicitly denote the assumption of $n\rightarrow\infty$ for the discussions in the rest of this paper.


The requirement of no information leakage to Eve in fact guarantees the highest possible decoding error probability at Eve.
As explained in~\cite[Remark~3.1]{Bloch_11}, consider that messages are uniformly taken from a size $K$ set $[1, 2, \cdots, K]$, and Eve minimizes her decoding error probability $P_{e}$ by performing maximum-likelihood decoding.
The condition of no information leakage ensures that Eve can only guess the original message, and the probability of error under maximum-likelihood  decoding is $P_{e}=\frac{K-1}{K}$. Therefore, from the decodability point of view, classical information-theoretic secrecy guarantees $P_e\ge\frac{K-1}{K}$.
Furthermore, when the entropy of the message is very large so that $K\rightarrow\infty$, classical information-theoretic secrecy actually guarantees that $P_e$ asymptotically goes to 1,
\begin{equation}\label{}
  \displaystyle\lim_{K\rightarrow\infty}P_e\ge\displaystyle\lim_{K\rightarrow\infty}\frac{K-1}{K}=1.
\end{equation}

In practice, the secrecy requirement on the decodability of messages at Eve can be generally written as  $P_{e}\ge\epsilon$ for some $\epsilon$.
Depending on the application, the value of $\epsilon$ ranges from 0 to 1, which falls outside the classical information-theoretic secrecy regime. 

\subsection{Partial Secrecy}
Partial secrecy is often quantified by the equivocation, which indicates the level at which Eve is confused.
In this paper, we specifically consider the fractional equivocation, which is defined as~\cite{Cheong_78}
\begin{equation}\label{eq:frac_equiv}
  \Delta=\frac{H(\M\mid \Z^n)}{H(\M)}.
\end{equation}
Note that evaluating security on the basis of equivocation is related to the conventional requirement on the decodability of messages at Eve\cite{wyner_75}. Although there is no one-to-one relation between the equivocation and the error probability, tight lower and upper bounds of the decoding error probability can be derived from the equivocation~\cite{Cover_06,Feder_94}.

When studying secrecy, we particularly want to ensure that the decoding error probability at the eavesdropper is larger than a certain level. Thus, it is desirable to have the decoding error probability at Eve lower bounded by the equivocation.
Still consider the general case where messages are uniformly taken from a size $K$ set $[1, 2, \cdots, K]$, which achieves the maximal entropy over an alphabet of size $K$. Then, the entropy of the message is given by $H(M)=\log_2(K)$. From Fano's inequality~\cite[Chapter~2.10]{Cover_06}, we have
\begin{equation}\label{}
  H(M \mid Z^n)\le h(P_e)+P_e\log_2(K),
\end{equation}
where $h(x)=-x\log_2(x)-(1-x)\log_2(1-x), ~0\le x\le1$.
This inequality can be weakened to
\begin{equation}\label{eq:PeDelta1}
  P_e\ge \frac{H(M \mid Z^n)-1}{\log_2(K)}=\Delta-\frac{1}{\log_2(K)}.
\end{equation}
When the entropy of the message is very large such that $K\rightarrow\infty$, we can further derive~\eqref{eq:PeDelta1} as
\begin{equation}\label{eq:PeboundbyDelta}
  \displaystyle\lim_{K\rightarrow\infty}P_e\ge\Delta-\displaystyle\lim_{K\rightarrow\infty}\frac{1}{\log_2(K)}=\Delta.
\end{equation}
Thus, $P_e$ is asymptotically lower bounded by $\Delta$.

\section{New Secrecy Metrics for Wireless Transmissions}\label{sec:content}

Consider the basic wiretap-channel system as introduced in the previous section. We now assume that the messages are transmitted over  quasi-static fading channels. Bob and Eve perfectly know their own CSI, but Eve's instantaneous CSI is not available at the legitimate side.
For wireless transmissions in such a system, classical information-theoretic secrecy is not always achievable, and the secrecy outage probability is commonly used to measure the secrecy performance.
From the classical information-theoretic secrecy perspective, the classical definition of secrecy outage probability treats the failure of achieving \textit{classical information-theoretic secrecy} as a secrecy outage.
Thus, the classical secrecy outage probability is applicable only for the system which has an extremely stringent requirement on Eve's decoding error probability, $\epsilon\rightarrow1$,
but cannot handle the general requirement on Eve's decoding error probability, $0<\epsilon\le1$.
In addition, the outage-based secrecy metric cannot evaluate how much or how fast the confidential information is leaked to Eve.

Unlike classcial secrecy outage probability, we study the secrecy performance of wireless communications from the partial secrecy perspective.
For wireless transmissions over fading channels, the fractional equivocation, $\Delta$, is a random quantity due to the fading properties of the channel.
Thus, we start from the derivation of $\Delta$ for a given fading realization.
The distribution of $\Delta$ can be obtained according to the distribution of the channel gains.
After that, three new secrecy metrics are proposed based on the distribution of $\Delta$.


%
%
\subsection{Fractional Equivocation for a Given Fading Realization}
A given fading realization of the wireless channel is equivalent to the (non-degraded) Gaussian wiretap channel\cite{Liang_08}.
The value of the fractional equivocation for the Gaussian wiretap channel actually depends on the coding and transmission strategies, and there is no general expression applicable for all scenarios. However, an upper bound on $\Delta$ can be easily derived following closely from~\cite[Theorem~1]{Cheong_78} and~\cite[Corollary~2]{Liang_08}.
The maximum achievable fractional equivocation for a given fading realization of the wireless channel is given by
\begin{equation}\label{eq:d_one}
  \Delta = \left\{ \begin{array}{lll} 1\;, & \text{if} \quad C_e\le C_b-R   \\
  (C_b-C_e)/R\;, & \text{if} \quad C_b-R<C_e<C_b \\
  0\;, & \text{if} \quad C_b\le C_e, \\
  \end{array}
  \right.
\end{equation}
where $C_b$ and $C_e$ denote Bob and Eve's channel capacities, respectively, and $R=\frac{H(M)}{n}$ denotes the secrecy rate for transmission.


\subsection{New Secrecy Metrics}\label{sec:content-Partial}
From \eqref{eq:d_one}, we note that $\Delta$ is a random quantity determined by the instantaneous channel gains and the transmission rate. Since the instantaneous knowledge of Eve's channel is unknown, we cannot directly characterize the instantaneous secrecy performance of the transmissions. Consequently, a meaningful system characterization relies on studying the distribution of $\Delta$,  which measures the long-term performance of the system with time-varying channel realizations.
In the following, we investigate the distribution of $\Delta$ from three aspects to propose three secrecy  metrics.


\vspace{1.5mm}
\subsubsection{Generalized Secrecy Outage Probability}
~\\
Extending the classical definition of secrecy outage probability, we propose a generalized definition of secrecy outage probability, given by
\begin{equation}\label{eq:Pout_G}
 p_{\text{out}}=\mathbb{P}\left(\Delta<\theta\right),
\end{equation}
where $\mathbb{P}\left(\cdot\right)$ denotes the probability measure and $0<\theta\le1$ denotes the minimum acceptable value of the fractional equivocation.

Since the fractional equivocation is related to the decoding error probability, the generalized secrecy outage probability is applicable for systems with different levels of secrecy requirements measured in terms of Eve's ability to decode the confidential messages (by choosing different values of $\theta$).
The classical secrecy outage probability is defined as $\mathbb{P}\left(\Delta<1\right)$, and hence is a special case of the new secrecy outage metric.
Apart from the discussion above, another way to understand the generalized secrecy outage probability can be described as follows. From \eqref{eq:frac_equiv}, the information leakage ratio to Eve can be written as $\frac{I(\M; \Z^n)}{H(\M)}\!=\!1-\Delta$. The information leakage ratio quantifies the percentage of transmitted confidential information leaked to the eavesdropper.
As such, the generalized secrecy outage probability, $p_{\text{out}}=\mathbb{P}\left(\Delta<\theta\right)=\mathbb{P}\left(1-\Delta>1-\theta\right)$, actually characterizes the probability that the information leakage ratio  is larger than a certain value, $1-\theta$. 

In fact,  we can also explain the generalized secrecy outage probability as an extension of partial secrecy in the Gaussian channel to the fading channel. Partial secrecy was originally proposed and investigated in the Gaussian channel in some of the pioneering studies of physical layer security, e.g.,~\cite{wyner_75,csiszar_78,Cheong_78}. It has also been adopted in evaluating the secrecy performance of finite-length codes in the Gaussian channel, e.g., \cite{Baldi_15_Performance,Wong_11_LDPC,Wong_11_SecretSharing}.
It is worth mentioning that a secrecy metric similar to the generalized secrecy outage probability was adopted in~\cite{Baldi_14_Secrecy}, which focused on analyzing the performance of finite-length codes in the fading channel.
In~\cite{Baldi_14_Secrecy}, a secrecy metric was adopted that quantifies the probability of Eve's decoding error being less than a given threshold, a result that was motivated by the fact that  finite-length codes cannot guarantee  Eve's decoding error rate will approach 1. The secrecy metric in~\cite{Baldi_14_Secrecy} is based on the partial secrecy metric adopted in~\cite{Wong_11_SecretSharing} for finite-length codes in the Gaussian channel. The fact that~\cite{Baldi_14_Secrecy} also adopts a partial secrecy metric further shows that classical secrecy outage probability has a severe limitation in evaluating the secrecy performance of wireless systems with finite-length codes.
\vspace{1.5mm}

\subsubsection{Average Fractional Equivocation -- Asymptotic Lower Bound on Eavesdropper's Decoding Error Probability}
~\\
Taking the average of the fractional equivocation, we can derive the (long-term) average value of the fractional equivocation, given by
\begin{equation}\label{eq:Dta_G}
  \Dta=\mathbb{E}\{\Delta\},
\end{equation}
where $\mathbb{E}\{\cdot\}$ denotes the expectation operation.
Note that the average fractional equivocation takes the average of the  values of fractional equivocation over all fading realizations. Since the fading varies slowly compared with one symbol time in quasi-static fading channels, it takes a relatively long time to experience a sufficient number of fading realizations during the transmissions. Thus, to be rigorous, we define $\Dta$ as the (\emph{long-term}) average fractional equivocation.
As discussed earlier in \eqref{eq:PeboundbyDelta}, Eve's decoding error probability for a given fading realization is asymptotically lower bounded by the fractional equivocation.  
Thus, the average fractional equivocation, $\Dta$, actually gives an asymptotic lower bound on the overall decoding error probability at Eve, i.e, $P_e\ge\Dta$.

\vspace{1.5mm}
\subsubsection{Average Information Leakage Rate}
~\\
With knowledge of message transmission rate $R=\frac{H(M)}n$, we can further derive the average information leakage rate, given by
\begin{equation}\label{eq:RL_G}
     R_L=\mathbb{E}\left\{\frac{I(\M; \Z^n)}{n}\right\}=\mathbb{E}\left\{(1-\Delta)R\right\}.
\end{equation}
The average information leakage rate tells how fast the information is leaked to the eavesdropper.
Note that the transmission rate $R$ cannot be simply taken out of the expectation in~\eqref{eq:RL_G}, since $R$ can be a variable parameter (e.g., adaptive-rate transmission) and its distribution may be correlated with the distribution of $\Delta$.
However, when a fixed-rate transmission scheme is adopted,
\eqref{eq:RL_G} can be simplified as
\begin{equation}\label{eq:RL_G_Fix}
  R_L=\mathbb{E}\left\{(1-\Delta)R\right\}=(1-\Dta)R.
\end{equation}
\begin{Remark}
The proposed secrecy metrics in this section, i.e.,~\eqref{eq:Pout_G},~\eqref{eq:Dta_G} and~\eqref{eq:RL_G}, are general and can be applied to evaluate the performance of any coding and transmission strategy under any system model (e.g., signal-antenna or multi-antenna systems).  A specific scenario is studied as an example in the next section, wherein the expressions for the proposed secrecy metrics are further derived in terms of transmission rates and channel statistics.
\end{Remark}

%


\section{Wireless Transmissions with Fixed-Rate Wiretap Codes: An Example}\label{sec:example}


\subsection{System Model}
We consider the system where a transmitter, Alice, wants to send confidential information to an intended receiver, Bob, in the present of an eavesdropper, Eve, over a quasi-static Rayleigh fading channel.
Alice, Bob and Eve are assumed to have a single antenna each.
The instantaneous channel capacities at Bob and Eve are given by
\begin{equation}\label{eq:Cb}
 C_b=\log_2(1+\vr_b)
\end{equation}
and
\begin{equation}\label{eq:Ce}
 C_e=\log_2(1+\vr_e),
\end{equation}
respectively, where $\vr_b$ and $\vr_e$ denote the instantaneous received signal-to-noise ratios (SNRs) at Bob and Eve, respectively.
The instantaneous received SNRs at Bob and Eve have exponential distributions, given by
\begin{equation}\label{eq:vr_b}
f_{\vr_b}(\vr_b)=\frac{1}{\vrba}\exp\left(-\frac{\vr_b}{\vrba}\right)
\end{equation}
and
\begin{equation}\label{eq:vr_e}
f_{\vr_e}(\vr_e)=\frac{1}{\vrea}\exp\left(-\frac{\vr_e}{\vrea}\right),
\end{equation}
respectively, where $\vrea$ and $\vrea$ denote the average received SNRs at Bob and Eve, respectively.

We consider the widely-adopted wiretap code \cite{wyner_75} for message transmissions. There are two rate parameters, namely, the codeword transmission rate, $R_b=\frac{H(X^n)}{n}$, and the confidential information rate, $R_s=\frac{H(M)}{n}$. 
A length $n$ wiretap code is constructed by generating $2^{nR_b}$ codewords $x^n(w,v)$, where $w=1,2,\cdots,2^{nR_s}$ and $v=1,2,\cdots,2^{n(R_b-R_s)}$.
For each message index $w$, we randomly select $v$ from $\left\{1,2,\cdots,2^{n(R_b-R_s)}\right\}$ with uniform probability and transmit the codeword $x^n(w,v)$.
In addition, we consider fixed-rate transmission,\footnote{Fixed-rate transmissions are often adopted to reduce system complexity. In practice, applications like video streaming in multimedia applications often require fixed-rate transmission.} where the transmission rates, i.e., $R_b$ and $R_s$, are fixed over time.

Bob and Eve are assumed to perfectly know their own channels. Hence, $C_b$ and $C_e$ are known at Bob and Eve, respectively.
Alice has statistical knowledge of Bob and Eve's channels, but does not know either Bob or Eve's instantaneous CSI.
We further assume that Bob provides a one-bit feedback about his channel quality to Alice in order to avoid unnecessary transmissions~\cite{Zhou_11,He_13_2}. The one-bit feedback enables an on-off transmission scheme to guarantee that the transmission takes place only when $R_b\le C_b$.  In addition, the on-off transmission scheme incurs a probability of transmission, given by
\begin{equation}\label{eq:ptx}
  \ptx=\mathbb{P}\left(R_b \le C_b\right)=\exp\left(-\frac{2^{R_b}-1}{\vrba}\right).
\end{equation}




\subsection{Secrecy Performance Evaluation}
To characterize the secrecy performance of wireless transmissions over the fading channel, we start from the investigation on a given fading realization of the channel.
\begin{Proposition}
{
For a given fading realization of the wireless channel, the maximum achievable fractional equivocation for the wiretap code with $R_b\le C_b$ and $R_s\le R_b$ is given by
\begin{equation}\label{eq:Cor1}
  \Delta = \left\{ \begin{array}{lll} 1\;, & \text{if} \quad C_e\le R_b-R_s   \\
  (R_b-C_e)/R_s\;, & \text{if} \quad R_b-R_s<C_e<R_b \\
  0\;, & \text{if} \quad R_b\le C_e. \\
  \end{array}
  \right.
\end{equation}
}
\end{Proposition}
\begin{IEEEproof}
The proof follows closely from~\cite[Corollary~2]{Liang_08} and the steps in~\cite[Section III]{Cheong_78} with $\frac{H(X^n)}{n}=R_b$.
\end{IEEEproof}
Note that $\Delta$ in~\eqref{eq:Cor1} actually gives an upper bound on the achievable fractional equivocation for the wiretap code, which is achieved by an ideal coding scheme with infinite codeword length.
It is worth mentioning that it is also of significant importance to obtain the lower bound of the fractional equivocation when investigating the performance of a specific code, e.g.,~\cite{Baldi_15_Performance,Wong_11_LDPC} which study finite-length LDPC codes. The secrecy performance guaranteed by a given code can be characterized by the lower bound on the fractional equivocation.


From \eqref{eq:Ce}, we can further derive~\eqref{eq:Cor1} as
\begin{equation}\label{eq:Cor2}
  \Delta = \left\{ \begin{array}{lll} 1\;,  & \text{if} \quad \vr_e \le 2^{R_b-R_s}-1   \\
  \frac{R_b-\log_2(1+\vr_e)}{R_s}\;, & \text{if} \quad 2^{R_b-R_s}-1 < \vr_e < 2^{R_b}\!-\!1 \\
  0\;, & \text{if} \quad 2^{R_b}-1 \le \vr_e. \\
  \end{array}
  \right.
\end{equation}

Now, we are ready to evaluate the secrecy performance of wireless transmissions over fading channels
from the distribution of $\Delta$, which can be derived according to the distribution of $\vr_e$ given in \eqref{eq:vr_e}.

\vspace{1.5mm}
\subsubsection{Generalized Secrecy Outage Probability}
~\\
The generalized secrecy outage probability is given by
\begin{eqnarray}\label{eq:result_1}
  p_{\text{out}}\!\!\!\!\!&=&\!\!\!\! \mathbb{P}(\Delta<\theta) \nonumber\\
  &=& \!\!\!\!\mathbb{P}\left(2^{R_b}-1\le\vr_e\right)+ \mathbb{P}\left(2^{R_b-R_s}-1<\vr_e<2^{R_b}-1\right)\nonumber\\
  &&\!\!\!\!\!\!\cdot~\mathbb{P}\left(\!\left.{\frac{R_b\!-\!\log_2(1\!+\!\vr_e)}{R_s}\!<\!\theta} \right\vert{\!2^{R_b\!-\!R_s}\!-\!1\!<\vr_e\!<2^{R_b}\!-\!1}\right) \nonumber\\
   &=&\!\!\!\! \exp\left(-\frac{2^{R_b-\theta R_s}-1}{\vrea}\right),
\end{eqnarray}
where $0<\theta\le1$.

For the extreme case of $\theta=1$, we have
\begin{equation}\label{eq:out_perfectfix}
   p_{\text{out}}(\theta=1) = \exp\left(-\frac{2^{R_b-R_s}-1}{\vrea}\right). 
\end{equation}
We note that \eqref{eq:out_perfectfix} is exactly the same as~\cite[Eq.~(8)]{Zhou_11}, which gives the classical secrecy outage probability of wireless transmissions with fixed-rate wiretap codes.

\vspace{1.5mm}
\subsubsection{Average Fractional Equivocation -- Asymptotic Lower Bound on Eavesdropper's Decoding Error Probability}
~\\
The average fractional equivocation is given by
\begin{eqnarray}\label{eq:ILRinExample}
  \Dta\!\!\!\!\!\!&=& \!\!\!\!\!\mathbb{E}\{\Delta\}\nonumber\\
\!\!\!\!\!\! &=& \!\!\!\!\! \int_{0}^{2^{R_b\!-\!R_s}\!-\!1}\!\!\!\!\!\!\!\!\!\!\!\!\!\!\!\!\!\!\!\! f_{\vr_e}(\vr_e)\mathrm{d}{\vr_e}\!+\!\int_{2^{R_b\!-\!R_s}\!-\!1}^{2^{R_b}\!-\!1}\!\!\left(
  \frac{R_b\!-\!\log_2(1\!+\!\vr_e)}{R_s}\right)\!f_{\vr_e}(\vr_e)\mathrm{d}{\vr_e} \nonumber \\
  \!\!\!\!\!\! &=&  \!\!\!\!\!1\!-\!\frac{1}{R_s\!\ln2}\exp\!\left(\!\frac{1}{\vrea}\!\right)\!
   \left(\!\mathrm{Ei}\!\left(\!-\frac{2^{R_b}}{\vrea}\!\right)\!-\!\mathrm{Ei}\!\left(\!-\frac{2^{R_b\!-\!R_s}}{\vrea}\right)\!\right),
\end{eqnarray}
where $\mathrm{Ei}\left(x\right)=\int^x_{-\infty}e^t/t~ \mathrm{d}{t}$ denotes the exponential integral function.
As mentioned before, the average fractional equivocation actually gives an asymptotic lower bound on the eavesdropper's decoding error probability.

\vspace{1.5mm}
\subsubsection{Average Information Leakage Rate}
~\\
Since a fixed-rate transmission scheme is adopted, the average information leakage rate can be derived from~\eqref{eq:RL_G_Fix},  given by
\begin{eqnarray}\label{eq:RL_fading_example}
\!\!\!\!\!\!\!\!\!\!R_L\!\!\!\!&=&\!\!\!\!(1-\Dta)R_s \nonumber\\
 \!\!\!\!&=&\!\!\!\!\frac{1}{\ln2}\exp\left(\!\frac{1}{\vrea}\!\right)\!\!
   \left(\!\mathrm{Ei}\left(\!-\frac{2^{R_b}}{\vrea}\right)\!-\!\mathrm{Ei}\left(\!-\frac{2^{R_b-R_s}}{\vrea}\!\right)\!\right),
\end{eqnarray}
which captures how fast on average information is leaked to Eve.
Note that the derivation of $R_L$ in \eqref{eq:RL_fading_example} does not depend on the probability of transmission $\ptx$, which indicates that $R_L$ actually characterizes how fast on average the information is leaked to the eavesdropper when a message transmission occurs.

\subsection{Numerical Results}\label{Sec:NumeralResultPart1}

\begin{figure}[!h]
\centering\vspace{-0mm}
\includegraphics[width=\columnwidth]{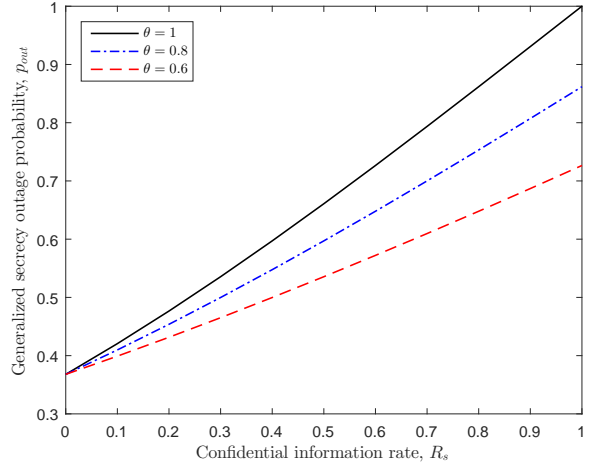}
\caption{Generalized secrecy outage probability versus confidential information rate. Results are shown for networks with different requirements on the fractional equivocation, $\theta=1, 0.8, 0.6$. The other parameters are $R_b=1$ and $\vrea=1$.}
\vspace{-0mm}  \label{fig:PoutVSRs}
\end{figure}

We first compare the generalized secrecy outage probabilities subject to different requirements on the fractional equivocation.
Figure~\ref{fig:PoutVSRs} plots $p_{\text{out}}$ versus $R_s$ with different values of $\theta$. Note that the case of $\theta=1$ represents classical secrecy outage probability.
As shown in the figure, for different levels of secrecy requirements measured in terms of the fractional equivocation or the decodability of messages at Eve, the transmission has different secrecy outage performance.
We find that the difference in the generalized secrecy outage probabilities increases as the confidential information rate increases.

\begin{figure}[!h]
\centering\vspace{-0mm}
\includegraphics[width=\columnwidth]{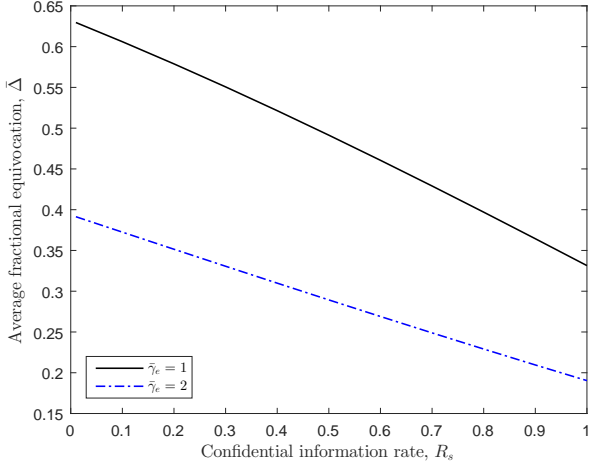}
\caption{Average fractional equivocation (asymptotic lower bound on the decoding error probability at Eve) versus confidential information rate. Results are shown for networks with different average received SNRs at Eve, $\vrea=1, 2$. The other parameter is $R_b=1$.}
\vspace{-0mm}  \label{fig:DeltavsRs}
\end{figure}

\begin{figure}[!h]
\centering\vspace{-0mm}
\includegraphics[width=\columnwidth]{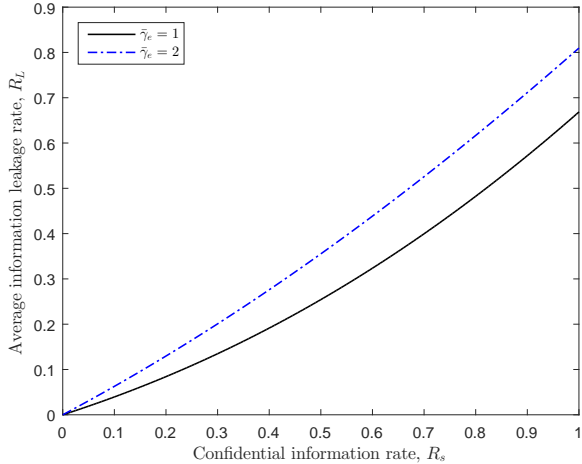}
\vspace{-0mm}
\caption{Average information leakage rate versus confidential information rate. Results are shown for networks with different average received SNRs at Eve, $\vrea=1, 2$. The other parameter is $R_b=1$.}
\vspace{-0mm}  \label{fig:RLVsRs}
\end{figure}

We then present the secrecy performance measured by the average fractional equivocation,
which gives an asymptotic lower bound on Eve's decoding error probability.
Figure~\ref{fig:DeltavsRs} plots $\Dta$ versus $R_s$. As shown in the figure, the average fractional equivocation decreases as the confidential information rate increases and/or the average received SNR at Eve increases. We note that the average fractional equivocation at Eve is not extremely high even when the confidential information rate is very small. We also note that the average fractional equivocation is non-zero even when the confidential information rate approaches the total transmission rate ($R_b=R_s$).
These observations indicate that the  quality of the wireless channel itself plays an important role in determining the secrecy performance of the wireless system.

Next, we illustrate the secrecy performance measured by the average information leakage rate.
Figure~\ref{fig:RLVsRs} plots $R_L$ versus $R_s$. As the figure shows, the average information leakage rate increases as the confidential information rate increases and/or the average received SNR at Eve increases. We note that $R_L$ does not reach $R_s$ even when $R_s$ goes to $R_b=1$. This implies that the information is not all leaked to the eavesdropper even when we use an ordinary code instead of the wiretap code for transmission.
This observation once again confirms that the wireless channel itself can provide a certain level of secrecy for the transmission.

\begin{figure*}[!htb]
  \subcaptionbox*{(a)}[.31\linewidth][c]{%
    \includegraphics[width=.36\linewidth]{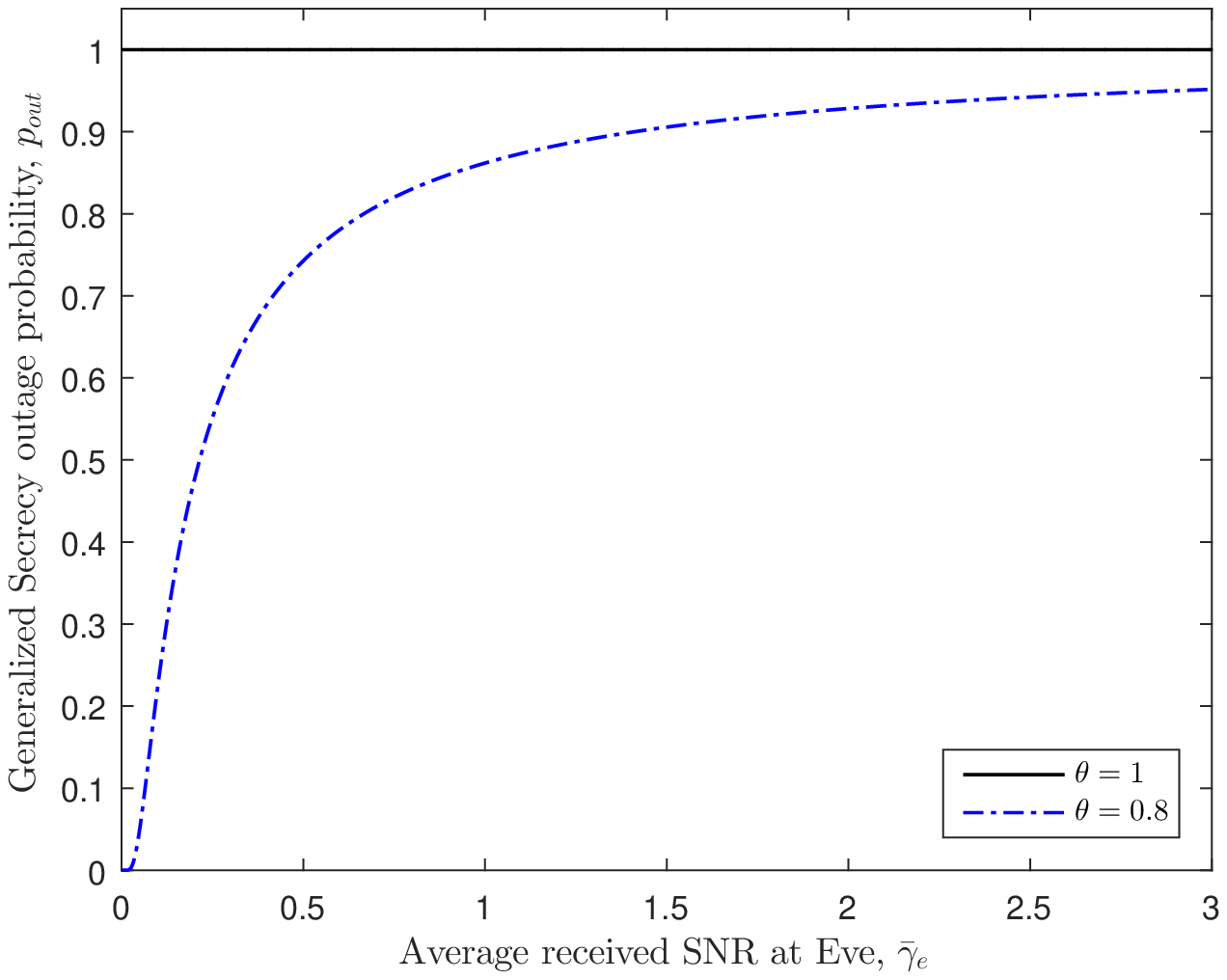}}\quad
  \subcaptionbox*{(b)}[.31\linewidth][c]{%
    \includegraphics[width=.36\linewidth]{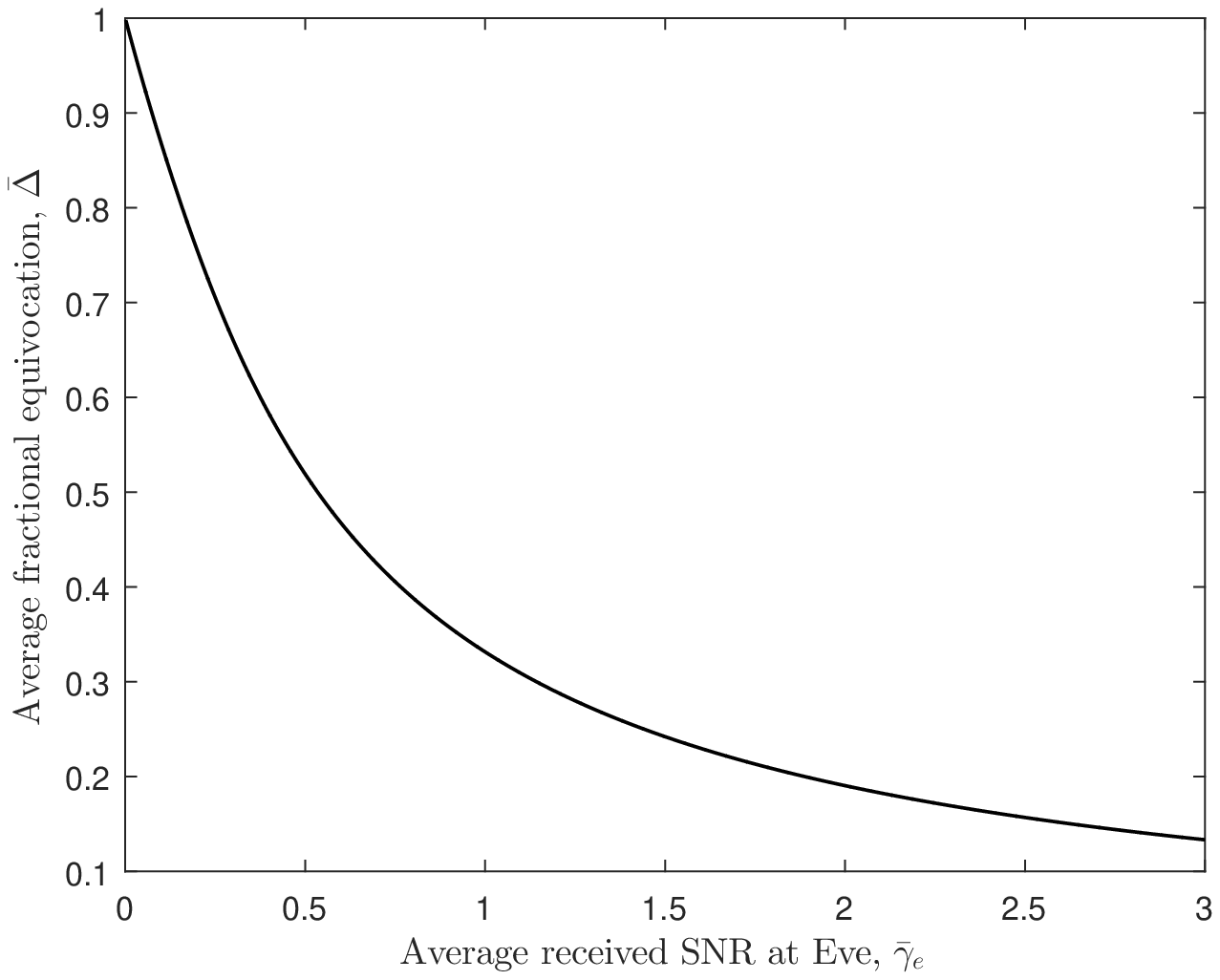}}\quad
  \subcaptionbox*{(c)}[.31\linewidth][c]{%
    \includegraphics[width=.36\linewidth]{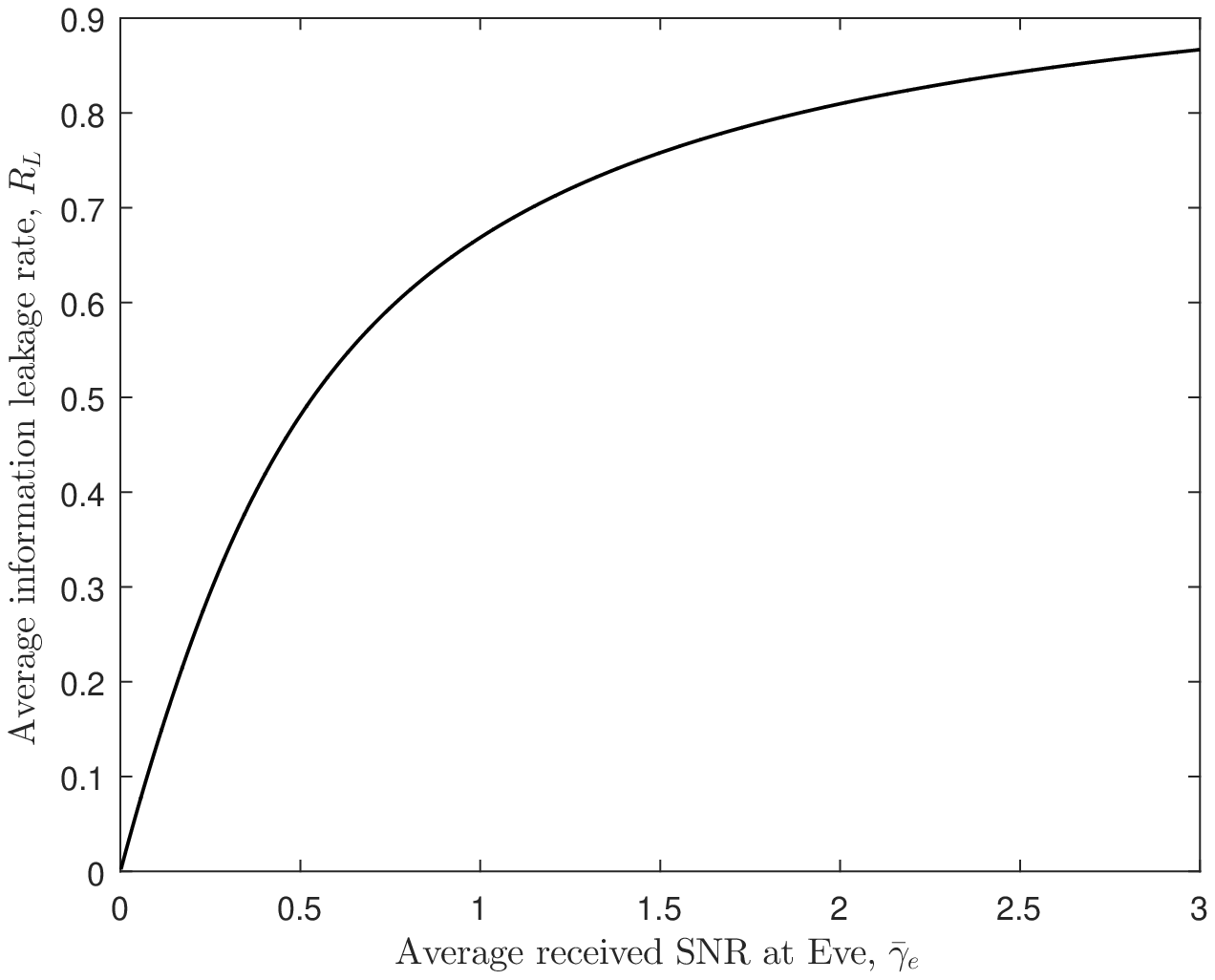}}
  \caption{Secrecy performance versus Eve's channel quality. Results are shown for the transmission with $R_b=R_s=1$. (a) Generalized secrecy outage probability versus average received SNR at Eve. (b) Average fractional equivocation versus average received SNR at Eve. (c) Average information leakage rate versus average received SNR at Eve.} \label{fig:F_s}
\end{figure*}


Finally, we show that the secrecy performance of wireless systems sometimes cannot be appropriately characterized by the classical secrecy outage probability, while on the other hand can be quantified by the newly purposed secrecy metrics.
In Figure~\ref{fig:F_s}, we evaluate the secrecy performance using classical secrecy outage probability and the newly proposed secrecy metrics for systems with different channel quality for Eve.
We consider an extreme case where the confidential information rate is the same as the total codeword rate, $R_b=R_s$. This is equivalent to using an ordinary code instead of the wiretap code for transmission.
As shown in Figure~\ref{fig:F_s}(a), the secrecy performance measured by the classical secrecy outage probability ($\theta=1$) is not related to Eve's channel condition, since it is always equal to 1.
However, we know that the decodability of messages at the receiver is related to the channel condition. Intuitively, with an improvement in Eve's channel quality, the probability of error at Eve should decrease, and the secrecy performance should become worse. Therefore, we see that the secrecy performance cannot be properly characterized by the classical secrecy outage probability.
In contrast, we find that the change of the secrecy performance with Eve's channel quality can be appropriately quantified by all three of the newly proposed secrecy metrics. In Figure~\ref{fig:F_s}(a), the generalized secrecy outage probability ($\theta=0.8$) increases as the average SNR at Eve increases.
In Figure~\ref{fig:F_s}(b), the average fractional equivocation decreases as the average SNR at Eve increases. In Figure~\ref{fig:F_s}(c), the average  information leakage rate increases as the average SNR at Eve increases.
This simple example of transmission with an ordinary code shows that the
newly proposed secrecy metrics are able to reveal information about the secrecy performance that cannot be captured by the classical secrecy outage probability.

\section{Impact on System Designs}\label{sec:Design}
In this section, we examine the significance of the newly proposed secrecy metrics from the perspective of a system designer, by answering the following questions:
\begin{enumerate}[Q1)]
  \item Do the newly proposed secrecy metrics lead to different system designs that optimize the secrecy performance, compared with the optimal design parameters minimizing the classical secrecy outage probability?
  \item Does applying the optimal transmission design based on the classical secrecy outage probability result in a large secrecy loss, if the actual system requires a low decodability at the eavesdropper or a low information leakage rate?
\end{enumerate}
As illustrated by the numerical results later in Section~\ref{sec:Numer:impact}, the answers to both Q1 and Q2 are yes, which shows that the newly proposed secrecy metrics have impact on the system design, and the impact is significant.
The fact that the answer to Q1 is yes implies that system designers cannot adopt the optimal design based on the classical secrecy outage probability to optimize the secrecy performance measured by the newly proposed secrecy metrics. The fact that the answer to Q2 is yes indicates that adopting the optimal design based on the classical secrecy outage probability would lead to a large
secrecy loss when the secrecy performance is measured by the newly proposed secrecy metrics.

\subsection{Problem Formulation}\label{Sec:ProForm}
We still consider the system with fixed-rate wiretap codes described in the previous section.
We optimize the secrecy performance of the wireless system subject to a throughput constraint $\eta>\Gamma$, where $\eta$ denotes the throughput of confidential message transmission and $\Gamma$ denotes its minimum required value.
The controllable parameters to design are the wiretap code rates $R_b$ and $R_s$.
Taking into account the probability of transmission given in \eqref{eq:ptx}, the throughput of the confidential message transmission is given by
\begin{equation}\label{eq:th_1}
  \eta=\ptx R_s=\exp\left(-\frac{2^{R_b}-1}{\vrba}\right)R_s.
\end{equation}

We specifically formulate three problems for the systems with different secrecy metrics as follows: 
\\
\emph{Problem~1:} Minimize the generalized secrecy outage probability
\begin{eqnarray}\label{eq:optprob1given}
  \min_{R_b, R_s} &&  p_{\text{out}}=\exp\left(-\frac{2^{R_b-\theta R_s}-1}{\vrea}\right), \\
  \text{s.t.}&&  \eta\ge\Gamma, R_b\ge R_s >0.
\end{eqnarray}
\emph{Problem~2:} Maximize the average fractional equivocation
\begin{eqnarray}\label{eq:optprob2given}
\!\!\!\!\!\! \!\!\!\!  \max_{R_b, R_s} \!\!\!\!\!&&\!\!\!\!\!\!  \Dta\!=\!1\!\!-\!\!\frac{1}{R_s\!\ln2}\!\exp\!\left(\!\!\frac{1}{\vrea}\!\right)\!
   \!\left(\!\!\mathrm{Ei}\!\left(\!\!-\frac{2^{R_b}}{\vrea}\!\!\right)\!\!-\!\!\mathrm{Ei}\!\left(\!\!-\frac{2^{R_b\!-\!R_s}}{\vrea}\right)\!\!\right)\!\!,~ \\
  \text{s.t.}\!\!\!\!\!\!&& \!\!\!\!   \eta\ge\Gamma, R_b\ge R_s >0.
\end{eqnarray}
\emph{Problem~3:} Minimize the average information leakage rate
\begin{eqnarray}\label{eq:optprob3given}
\!\!\!\!\!\! \!\!\!\! \min_{R_b, R_s} \!\!\!\!\!\!&&\!\!\!\!\!\! R_L\!\!=\!\!\frac{1}{\ln2}\exp\left(\!\frac{1}{\vrea}\!\right)\!\!
   \left(\!\mathrm{Ei}\left(\!-\frac{2^{R_b}}{\vrea}\right)\!-\!\mathrm{Ei}\left(\!-\frac{2^{R_b\!-\!R_s}}{\vrea}\!\right)\!\!\right)\!\!,~ \\
  \text{s.t.}\!\!\!\!\!\!&& \!\!\!\!  \eta\ge\Gamma, R_b\ge R_s >0.
\end{eqnarray}

\subsection{Feasibility of the Constraint}
The required throughput constraint is not feasible when $\Gamma$ is larger than the maximum achievable throughput for $R_b\ge R_s>0$.
We find that the three problems have the same feasible constraint region, which is given by the following proposition.
\begin{Proposition}\label{pro:feasible}
The feasible range of the throughput constraint is given by
\begin{equation}\label{eq:etamaxfeasible}
  0\le\Gamma\le\frac{W_0(\vrba)}{\ln2}\exp\left(-\frac{2^{\frac{W_0(\vrba)}{\ln2}}-1}{\vrba}\right),
\end{equation}
where $W_0(\cdot)$ denotes the principal branch of the Lambert W function.
\end{Proposition}
\begin{IEEEproof}
See Appendix~\ref{AP:feasible}.
\end{IEEEproof}

\subsection{Optimal Rate Parameters}\label{sec:Op_solutions}
We denote $\Rsmin$ and $\Rsmax$ as the solutions of $x$ to
$\exp\left(-\frac{2^{x}-1}{\vrba}\right)x=\Gamma$
with $\Rsmin<\Rsmax$. The optimal solutions to Problems 1, 2 and 3 are summarized in Propositions \ref{pro:Solution1}, \ref{pro:Solution2} and \ref{pro:Solution3}, respectively, as follows.
\begin{Proposition}\label{pro:Solution1}
The optimal rate parameters minimizing the generalized secrecy outage probability are given as follows:
\begin{equation}\label{eq:R_bo1}
  R_{b1}^*=\log_2\left(1-\vrba\ln\frac{\Gamma}{R_{s1}^*}\right)
\end{equation}
and
\begin{equation}\label{eq:R_so1}
  R_{s1}^*=\left\{ \begin{array}{lll}
   \Rsmin\;, &\mbox{if~} \Rsmin>R_{so}     \\
   R_{so}\;, &\mbox{if~} \Rsmin\le R_{so}\le \Rsmax\\
   \Rsmax\;,  &\mbox{if~} \Rsmax<R_{so},    \end{array}
  \right.
\end{equation}
where $R_{so}$ is the solution of $x$ to
\begin{equation}\label{}
  \theta=\frac{\vrba}{x\ln(2)\left(1-\vrba\ln\left(\frac{\Gamma}{x}\right)\right)}.
\end{equation}
\end{Proposition}
\begin{IEEEproof}
See Appendix~\ref{AP:S_P1}.
\end{IEEEproof}

\begin{Proposition}\label{pro:Solution2}
The optimal rate parameters maximizing the average fractional equivocation are given as follows:
\begin{equation}\label{eq:R_bo2}
  R_{b2}^*=\log_2\left(1-\vrba\ln\frac{\Gamma}{R_{s2}^*}\right)
\end{equation}
and $R_{s2}^*$ is obtained by numerically solving the following problem:
\begin{eqnarray}\label{eq:opt_Rs2inProp5}
  \min_{x} \!\!\!\!&& \!\!\!\! \frac{1}{x}\!
   \left(\!\mathrm{Ei}\left(\!-\frac{1\!-\!\vrba\ln\frac{\Gamma}{x}}{\vrea}\!\right)\!-\!\mathrm{Ei}\left(\!-\frac{1-\vrba\ln\frac{\Gamma}{x}}{\vrea2^{x}}\right)\!\right)\!, \\
  \text{s.t.}\!\!\!\!&& \!\!\!\!  \Rsmin\le x\le\Rsmax.
\end{eqnarray}
\end{Proposition}
\begin{IEEEproof}
See Appendix~\ref{AP:S_P2}.
\end{IEEEproof}

\begin{Proposition}\label{pro:Solution3}
The optimal rate parameters minimizing the average information leakage rate  are given as follows:
\begin{equation}\label{}
  R_{b3}^*=\log_2\left(1-\vrba\ln\frac{\Gamma}{R_{s3}^*}\right)
\end{equation}
and $R_{s3}^*$ is obtained by numerically solving the following problem:
\begin{eqnarray}\label{}
  \min_{x} &&
 \mathrm{Ei}\left(-\frac{1-\vrba\ln\frac{\Gamma}{x}}{\vrea}\!\right)-\mathrm{Ei}\left(-\frac{1-\vrba\ln\frac{\Gamma}{x}}{\vrea2^{x}}\right), \\
  \text{s.t.}&&   \Rsmin\le x\le\Rsmax.
\end{eqnarray}
\end{Proposition}
\begin{IEEEproof}
The proof follows closely from the proof of Proposition~\ref{pro:Solution2} in Appendix~\ref{AP:S_P2}.
\end{IEEEproof}

\begin{Remark}
The numerical optimization problems for obtaining $R_{s2}^*$ and $R_{s3}^*$ in Propositions~\ref{pro:Solution2} and~\ref{pro:Solution3} can be easily solved by either a simple brute-force search or techniques like the golden section search~\cite{Kiefer_53_sequential}.
\end{Remark}

\subsection{Numerical Results}\label{sec:Numer:impact}
\begin{figure}[!htb]
\centering\vspace{-0mm}
\includegraphics[width=\columnwidth]{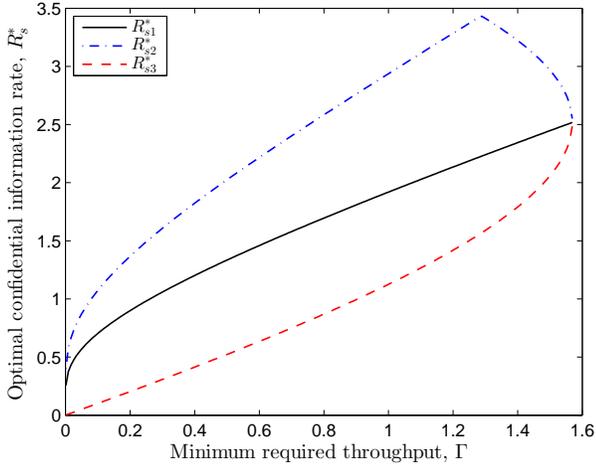}
\vspace{-0mm}
\caption{For different secrecy metrics: optimal confidential information rate  versus minimum required throughput. The other parameters are $\theta=1$, $\vrba=10$ dB and $\vrea=10$ dB.}
\vspace{-0mm}  \label{fig:ComThreeS}
\end{figure}

\begin{figure}[!htb]
\centering\vspace{-0mm}
\includegraphics[width=\columnwidth]{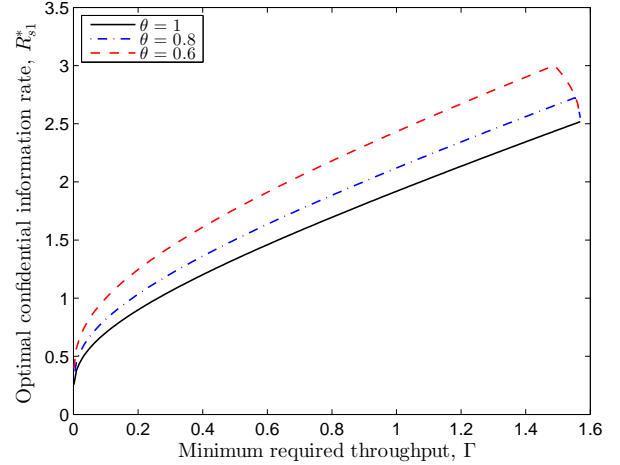}
\vspace{-0mm}
\caption{For generalized secrecy outage probability: optimal confidential information rate versus minimum required throughput. Results are shown for networks with different requirements on the fractional equivocation, $\theta=1, 0.8, 0.6$. The other parameters are $\vrba=10$ dB and $\vrea=10$ dB.}
\vspace{-0mm}  \label{fig:G_PoutSol}
\end{figure}

\begin{figure}[!htb]
\centering\vspace{-0mm}
\includegraphics[width=\columnwidth]{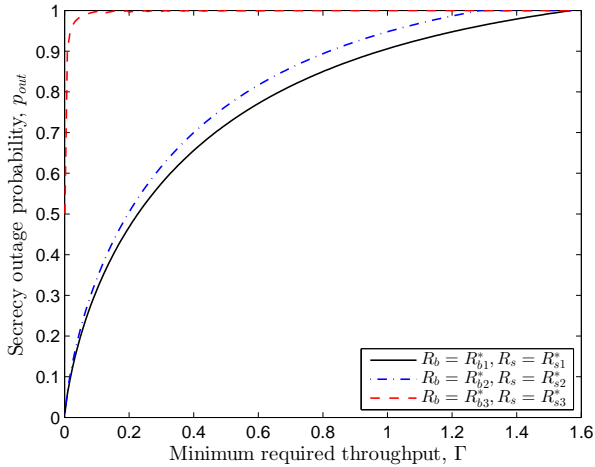}
\vspace{-0mm}
\caption{Secrecy outage probability versus minimum required throughput. The other parameters are $\theta=1$, $\vrba=10$ dB and $\vrea=10$ dB.}
\vspace{-0mm}  \label{fig:PoutSol}
\end{figure}

\begin{figure}[!htb]
\centering\vspace{-0mm}
\includegraphics[width=\columnwidth]{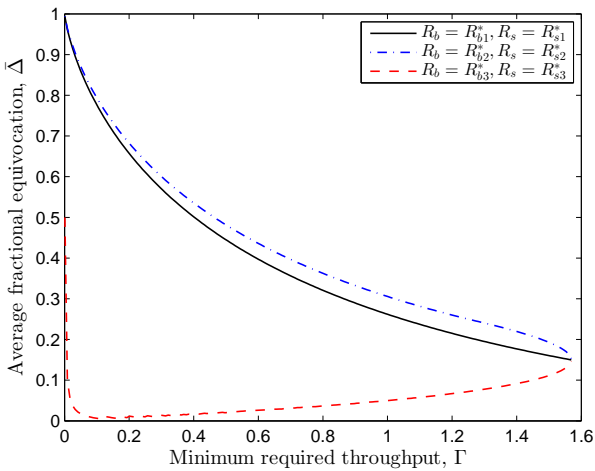}
\vspace{-0mm}
\caption{Average fractional equivocation (asymptotic lower bound on the decoding error probability at Eve) versus minimum required throughput. The other parameters are $\theta=1$, $\vrba=10$ dB and $\vrea=10$ dB.}
\vspace{-0mm}  \label{fig:DtaSol}
\end{figure}

\begin{figure}[!htb]
\centering\vspace{-0mm}
\includegraphics[width=\columnwidth]{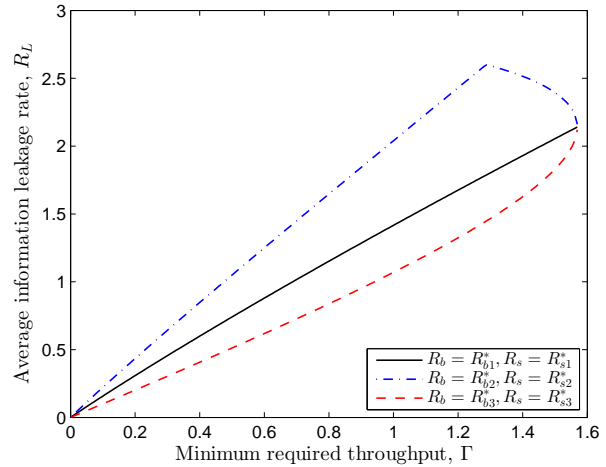}
\vspace{-0mm}
\caption{Average information leakage rate versus minimum required throughput. The other parameters are $\theta=1$, $\vrba=10$ dB and $\vrea=10$ dB.}
\vspace{-0mm}  \label{fig:RLSol}
\end{figure}

In this subsection, we present numerical results for a wireless system with $\vrba=10$~dB and $\vrea=10$~dB to demonstrate the impact of the new secrecy metrics on system designs. The feasible range of the throughput constraint is $0\le\Gamma\le 1.569$, which is obtained by Proposition~\ref{pro:feasible}.
Specifically, we can find the answer to Q1 by examining Figures~\ref{fig:ComThreeS} and~\ref{fig:G_PoutSol} and we can find the answer to Q2 by examining Figures~\ref{fig:PoutSol}, \ref{fig:DtaSol} and \ref{fig:RLSol}.

We first compare the transmission rates that optimize the secrecy performance of the system measured by different secrecy metrics.
Figure~\ref{fig:ComThreeS} plots the optimal confidential information rate $R_s^*$ versus the throughput constraint $\Gamma$.
The values of $R_{s1}^*$, $R_{s2}^*$ and $R_{s3}^*$ are obtained by Propositions~\ref{pro:Solution1},~\ref{pro:Solution2} and~\ref{pro:Solution3}, respectively. The optimal codeword transmission rate $R_b^*$ is not shown in the figure, since the optimal codeword transmission rate is equal to $R_b^*=\log_2\left(1-\vrba\ln\frac{\Gamma}{R_{s}^*}\right)$ for all three problems, and the differences between $R_{b1}^*$, $R_{b2}^*$ and $R_{b3}^*$ are determined by the differences between $R_{s1}^*$, $R_{s2}^*$ and $R_{s3}^*$.   
As depicted in the figure, the values of $R_{s1}^*$, $R_{s2}^*$ and $R_{s3}^*$  are clearly different from each other.
We note that $R_{s1}^*= R_{s2}^* = R_{s3}^*$ if and only if the throughput constraint is very stringent, in which case the transmission rates are totally determined by the throughput constraint.
The observations above illustrate that the optimal transmission designs are very different when we use different secrecy metrics to evaluate secrecy performance.

Next, we focus on  the optimal transmission rates that minimize the generalized secrecy outage probabilities subject to different requirements on the fractional equivocation. Figure~\ref{fig:G_PoutSol} plots $R_{s1}^*$ versus $\Gamma$ for different values of $\theta$. As shown in the figure, the optimal transmission rates minimizing the secrecy outage probability are different if the required values of $\theta$ are different. We find that the optimal confidential information rate $R_{s1}^*$ increases as the level of required fractional equivocation $\theta$ decreases.
The observations from Figures~\ref{fig:ComThreeS} and~\ref{fig:G_PoutSol} confirm that the answer to Q1 is yes: the newly proposed secrecy metrics lead to very different system design choices that optimize the secrecy performance.

In the following, we answer the second question listed at the beginning of this section
using Figures~\ref{fig:PoutSol}, \ref{fig:DtaSol} and \ref{fig:RLSol}.
From the analytical results, we have obtained three different solutions of the optimal design parameters: $\left(R_{b1}^*, R_{s1}^*\right)$ is optimal for minimizing the generalized secrecy outage probability; $\left(R_{b2}^*, R_{s2}^*\right)$ is optimal for maximizing the average fractional equivocation; $\left(R_{b3}^*, R_{s3}^*\right)$ is optimal for minimizing the average information leakage rate. We collectively consider all three design solutions and study their performance for all three secrecy metrics.
Specifically, Figure~\ref{fig:PoutSol} plots $p_{\text{out}}$, Figure~\ref{fig:DtaSol} plots $\Dta$, and Figure~\ref{fig:RLSol} plots $R_L$ achieved by the different design strategies.
As shown in the figures, transmission with $R_{b1}^*$ and $R_{s1}^*$ minimizes the secrecy outage probability, but leads to a considerable loss if the practical secrecy requirement is to ensure a high fractional equivocation (decoding error probability at Eve) or a low information leakage rate.
Similarly, transmission with $R_{b2}^*$ and $R_{s2}^*$ maximizes the average fractional equivocation, but incurs a considerable loss if the practical secrecy requirement is to have a low secrecy outage probability or a low information leakage rate. Finally, transmission with $R_{b3}^*$ and $R_{s3}^*$ minimizes the average information leakage rate, but incurs a large loss if the practical secrecy requirement is to maintain a low secrecy outage probability or a high fractional equivocation.
The observations from Figures~\ref{fig:PoutSol},~\ref{fig:DtaSol} and~\ref{fig:RLSol} show that it is important to design the system with the appropriate secrecy metric. It is also confirmed that the answer to Q2 is yes: applying the transmission design based on the classical secrecy outage probability can result in a large secrecy loss if the actual system requires a low decodability at the eavesdropper or a low information leakage rate.


\section{Conclusion and Future Work}\label{sec:Conc}
To address the practical limitations of using classical secrecy outage probability as a metric for secrecy, we proposed three new metrics for physical layer security over quasi-static fading channels.
Specifically, the generalized secrecy outage probability establishes a link between the concept of secrecy outage and the decodability of messages at the eavesdropper. The asymptotic lower bound on the eavesdropper's decoding error probability provides a direct error-probability-based secrecy metric. The average information leakage rate characterizes how fast the confidential information is leaked to the eavesdropper when classical information-theoretic secrecy is not achieved.
We evaluated the performance of an example wireless system with fixed-rate wiretap codes using the proposed secrecy metrics.
We showed that the new secrecy metrics provide a more comprehensive understanding of physical layer security over fading channels.
We also found that the new secrecy metrics can give insights on the secrecy performance of wireless transmissions that sometimes cannot be captured by classical secrecy outage probability. Furthermore, we examined the significance of the newly proposed secrecy metrics from the perspective of a system designer.  We found that applying the optimal transmission design minimizing the classical secrecy outage probability can result in a large secrecy loss, if the actual system requires a low decodability at the eavesdropper or a low information leakage rate.
The new secrecy metrics enable appropriate transmission designs for systems with different secrecy requirements.
We hope that this work can help bridge the gap between theory and practice in physical layer security by inspiring more future studies adopting and building on the newly proposed secrecy metrics.
Besides, as mentioned previously in Section~\ref{Sec:IA}, it is of importance to investigate secrecy metrics for wireless systems with finite-length coding schemes, since the classical information-theoretic secrecy cannot be achieved by finite-length codes. While the secrecy metrics proposed in this work did not focus on the finite-length coding schemes, it is also a very interesting future research direction to investigate appropriate secrecy metrics specifically for wireless systems with finite-length codes.


\appendices
\section{Proof of Proposition~\ref{pro:feasible}}\label{AP:feasible}
To determine the maximum achievable secrecy throughput, we first obtain the optimal rate parameters that maximize the secrecy throughput.
The problem is formulated as
\begin{eqnarray}\label{eq:optprobgiven}
  \max_{R_b, R_s} &&  \eta=\exp\left(-\frac{2^{R_b}-1}{\vrba}\right)R_s, \\
  \text{s.t.}&&   R_b\ge R_s >0.
\end{eqnarray}
Given any $R_s$, we find that $\partial\eta/\partial R_b$ is always less than 0. Hence given any $R_s$, it is wise to have the minimum $R_b$, i.e., $R_b=R_s$, for maximizing $\eta$.
Then, the problem changes to
\begin{eqnarray}\label{eq:optprobgiven2}
  \max_{R_s} &&  \eta\left(R_b=R_s\right)=\exp\left(-\frac{2^{R_s}-1}{\vrba}\right)R_s, \label{eq:etaoptfeasible2}\\
  \text{s.t.}&&     R_s>0.
\end{eqnarray}
Taking the first order derivative of $\eta\left(R_b=R_s\right)$ with respect to $R_s$, we have
\begin{equation}\label{}
  \frac{\partial \eta\left(R_b\!=\!R_s\right)}{\partial R_s}\!=\!\exp\left(\!-\frac{2^{R_s}-1}{\vrba}\right)\left(1-\frac{2^{R_s}R_s\ln2}{\vrba}\right).
\end{equation}
By solving for $R_s$ in $\frac{\partial \eta\left(R_b=R_s\right)}{\partial R_s}=0$, we obtain the optimal value of $R_s$ that maximizes $\eta$, which is given by
\begin{equation}\label{eq:Rbsfeasible}
  R_s^\diamond=\frac{W_0(\vrba)}{\ln2}.
\end{equation}
Finally, substituting $R_s=R_s^\diamond$ into \eqref{eq:etaoptfeasible2} completes the proof.

\section{Proof of Proposition~\ref{pro:Solution1}}\label{AP:S_P1}
As analyzed in Appendix~\ref{AP:feasible}, given any $R_s$, it is wise to have the minimum $R_b$, i.e., $R_b=R_s$, for maximizing $\eta$. Hence, we can obtain the feasible range of $R_s$ for satisfying the throughput constraint
by solving $R_s$ in the equation  $\eta\left(R_b=R_s\right)=\Gamma$. The feasible range is given by $\Rsmin\le R_s\le\Rsmax$.

From $ p_{\text{out}}=\exp\left(-\frac{2^{R_b-\theta R_s}-1}{\vrea}\right)$, we find that minimizing $\pout$ is equivalent to maximizing
\begin{equation}\label{eq:O1ex}
  O_1=R_b-\theta R_s.
\end{equation}
To minimize $ O_1$ in \eqref{eq:O1ex}, it is wise to have the maximum $R_b$ while satisfying the throughput constraint, for any given $R_s$.
From $\eta=\exp\left(-\frac{2^{R_b}-1}{\vrba}\right)R_s\ge\Gamma$, we have
\begin{equation}\label{}
  R_b\le\log_2\left(1-\vrba\ln\frac{\Gamma}{R_s}\right).
\end{equation}
Hence, we obtain $R_{b1}^*$ as in~\eqref{eq:R_bo1}.
Then, we can rewrite the optimization problem as
\begin{eqnarray}\label{}
  \max_{R_s} &&  \log_2\left(1-\vrba\ln\frac{\Gamma}{R_s}\right)-\theta R_s, \\
  \text{s.t.}&&  \Rsmin\le R_s \le \Rsmax.
\end{eqnarray}
Finally, by solving for $R_s$ in the equation $\frac{\partial O_1}{\partial R_s}=0$ and considering the feasible range of $R_s$, we obtain $R_{s1}^*$ as in~\eqref{eq:R_so1}. This completes the proof.

\section{Proof of Proposition~\ref{pro:Solution2}}\label{AP:S_P2}
The feasible range of $R_s$ for satisfying the throughput constraint is given by $\Rsmin\le R_s\le\Rsmax$.
From $\Dta=1-\frac{1}{R_s\ln2}\exp\left(\frac{1}{\vrea}\right)
   \left(\mathrm{Ei}\left(-\frac{2^{R_b}}{\vrea}\right)-\mathrm{Ei}\left(-\frac{2^{R_b-R_s}}{\vrea}\right)\right) $, we find that maximizing $\Dta$ is equivalent to minimizing
\begin{equation}\label{eq:O2ex}
  O_2=\frac{1}{R_s}
   \left(\mathrm{Ei}\left(-\frac{2^{R_b}}{\vrea}\right)-\mathrm{Ei}\left(-\frac{2^{R_b-R_s}}{\vrea}\right)\right).
\end{equation}
Given any $R_s$, we have
\begin{equation}\label{}
  \frac{\partial O_2}{\partial R_b}=\frac{\ln(2)}{R_s}\left(\exp\left(-\frac{2^{R_b}}{\vrea}\right)-\exp\left(-\frac{2^{R_b-R_s}}{\vrea}\right)\right)<0.
\end{equation}
Hence given any $R_s$, it is wise to have the maximum $R_b$ while satisfying the throughput
constraint to minimize $O_2$ in~\eqref{eq:O2ex}. Hence, we obtain  $R_{b2}^*$ as in~\eqref{eq:R_bo2}. Then, we  rewrite the optimization problem as~\eqref{eq:opt_Rs2inProp5}.
We find that the closed-form solution of $R_{s2}^*$ is mathematically intractable. We can obtain $R_{s2}^*$ by numerically solving the problem. This completes the proof.

\bibliographystyle{IEEEtran}

\balance

\end{document}